\documentclass[aps,prl,twocolumn,superscriptaddress,amsmath]{revtex4-1}

\usepackage{graphicx}

\begin{document}

% Use the \preprint command to place your local institutional report
% number in the upper righthand corner of the title page in preprint mode.
% Multiple \preprint commands are allowed.
% Use the 'preprintnumbers' class option to override journal defaults
% to display numbers if necessary
%\preprint{}

%Title of paper
\title{Observation of Andreev bound states at spin-active interfaces}

% repeat the \author .. \affiliation  etc. as needed
% \email, \thanks, \homepage, \altaffiliation all apply to the current
% author. Explanatory text should go in the []'s, actual e-mail
% address or url should go in the {}'s for \email and \homepage.
% Please use the appropriate macro foreach each type of information

% \affiliation command applies to all authors since the last
% \affiliation command. The \affiliation command should follow the
% other information
% \affiliation can be followed by \email, \homepage, \thanks as well.
\author{F. H\"ubler}
\affiliation{Institut f\"ur Nanotechnologie, Karlsruher Institut f\"ur Technologie, Karlsruhe, Germany}
\affiliation{Center for Functional Nanostructures, Karlsruher Institut f\"ur Technologie, Karlsruhe, Germany}
\affiliation{Institut f\"ur Festk\"orperphysik, Karlsruher Institut f\"ur Technologie, Karlsruhe, Germany}
\author{M. J. Wolf}
\affiliation{Institut f\"ur Nanotechnologie, Karlsruher Institut f\"ur Technologie, Karlsruhe, Germany}
\author{T. Scherer}
\author{D. Wang}
\affiliation{Institut f\"ur Nanotechnologie, Karlsruher Institut f\"ur Technologie, Karlsruhe, Germany}
\affiliation{Karlsruhe Nano Micro Facility, Karlsruher Institut f\"ur Technologie, Karlsruhe, Germany}
\author{D. Beckmann}
\email[e-mail address: ]{detlef.beckmann@kit.edu}
\affiliation{Institut f\"ur Nanotechnologie, Karlsruher Institut f\"ur Technologie, Karlsruhe, Germany}
\affiliation{Center for Functional Nanostructures, Karlsruher Institut f\"ur Technologie, Karlsruhe, Germany}
\author{H. v. L\"ohneysen}
\affiliation{Center for Functional Nanostructures, Karlsruher Institut f\"ur Technologie, Karlsruhe, Germany}
\affiliation{Institut f\"ur Festk\"orperphysik, Karlsruher Institut f\"ur Technologie, Karlsruhe, Germany}
\affiliation{Physikalisches Institut, Karlsruher Institut f\"ur Technologie, Karlsruhe, Germany}
%Collaboration name if desired (requires use of superscriptaddress
%option in \documentclass). \noaffiliation is required (may also be
%used with the \author command).
%\collaboration can be followed by \email, \homepage, \thanks as well.
%\collaboration{}
%\noaffiliation

\date{\today}

\begin{abstract}
We report on high-resolution differential conductance experiments on nanoscale superconductor/ferromagnet tunnel junctions with ultra-thin oxide tunnel barriers. We observe subgap conductance features which are symmetric with respect to bias, and shift according to the Zeeman energy with an applied magnetic field. These features can be explained by resonant transport via Andreev bound states induced by spin-active scattering at the interface. From the energy and the Zeeman shift of the bound states, both the magnitude and sign of the spin-dependent interfacial phase shifts between spin-up and spin-down electrons can be determined. These results contribute to the microscopic insight into the triplet proximity effect at spin-active interfaces.
\end{abstract}

% insert suggested PACS numbers in braces on next line
\pacs{72.25.Mk, 74.45.+c, 73.63.-b, 85.25.-j}
% 72.25.Mk 	Spin transport through interfaces 
% 73.63.-b 	Electronic transport in nanoscale materials and structures
% 74.45.+c 	Proximity effects; Andreev reflection; SN and SNS junctions
% 85.25.-j 	Superconducting devices
% insert suggested keywords - APS authors don't need to do this
%\keywords{}

%\maketitle must follow title, authors, abstract, \pacs, and \keywords
\maketitle

% body of paper here - Use proper section commands
% References should be done using the \cite, \ref, and \label commands
In superconductors electrons are bound in Cooper pairs, usually in a singlet state, i.e., with opposite spin. Recently, flow of a supercurrent through fully spin-polarized chromium dioxide between two singlet superconductors has been reported \cite{keizer2006} indicating Cooper pairs of equal spin, thus corresponding to a long-range triplet proximity effect \cite{bergeret2001,*bergeret2005,*buzdin2005}. This phenomenon is believed to be due to a combination of spin-dependent interfacial phase shifts \cite{millis1988} of the electron wave-functions in conjunction with spin-flip scattering \cite{eschrig2008}. To date, most experimental evidence for the triplet proximity effect and spin-active scattering is based on the observation of the thickness dependence of the Josephson current \cite{kontos2002,*cottet2005,*khaire2010,*robinson2010,*anwar2010}, but there is an increasing interest in obtaining direct spectroscopic evidence \cite{loefwander2010,*usman2011}. In this Letter, we report on the observation of Andreev bound states at spin-active interfaces, allowing a precise determination of spin-dependent phase shifts.

Andreev scattering at interfaces is a powerful probe of the superconducting order-parameter symmetry \cite{deutscher2005}. Zero-energy Andreev bound states have been observed in both singlet {\it d}-wave \cite{tanaka1995,*kashiwaya1995} and triplet {\it p}-wave \cite{yamashiro1997,*laube2000} superconductors. Microscopically, these bound states occur due to wave-mechanical phase shifts between the electron and the hole involved in Andreev reflection. In the case of unconventional bulk superconductors, a phase shift of $\pi$ may be induced by sign reversals of the bulk order parameter, yielding a zero-energy bound state. Since the electron and the hole involved in Andreev reflection reside in opposite spin bands, spin-active interfaces also introduce a phase shift $\theta_\mathrm{s}$ into the Andreev process, but in this case any value between $-\pi$ and $\pi$ can occur. Consequently, the surface bound states can have arbitrary subgap energy \cite{zhao2004}. Andreev bound states at spin-active interfaces are predicted to induce characteristic features in the density of states \cite{zhao2004,bergeret2005b}, in the local \cite{zhao2004,bergeret2005b,grein2010} and non-local \cite{kalenkov2007,*metalidis2010} conductance as well as in the noise spectrum \cite{cottet2008}. 

\begin{figure}
\includegraphics[width=\columnwidth]{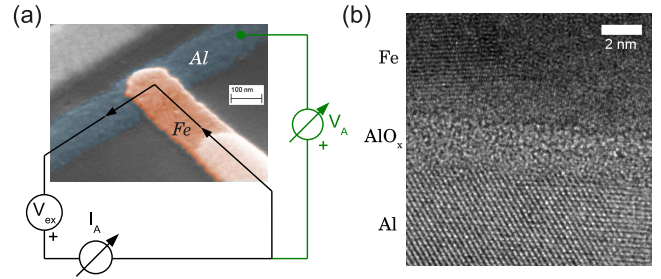}
\caption{\label{fig_scheme}(Color online)
(a) Scanning electron microscopy image of one of our samples, together with the measurement scheme.
(b) High-resolution transmission electron microscopy cross section of a reference structure, showing the atomic structure of the tunnel barrier.}
\end{figure}

In particular, the bound states give rise to a spin-dependent subgap density of states, with resonances at the characteristic energy \cite{zhao2004}
\begin{equation}
\epsilon_\pm=\pm\mathrm{sign}(\theta_\mathrm{s})\Delta\cos(\theta_\mathrm{s}/2), \label{equ_epsilon}
\end{equation}
for spin up ($\epsilon_+$) and down ($\epsilon_-$) quasiparticles, where $\Delta$ is the pair potential. In the differential conductance, the bound states show up as double-peak features due to resonant Andreev reflection at voltages $eV=-\epsilon_\pm$. The very simple relation between bound-state energy and phase shift allows an unambiguous and precise determination of $\theta_\mathrm{s}$. For small $\theta_\mathrm{s}$, however, the features are expected to be very close to $\Delta$, and therefore difficult to distinguish from the ordinary gap anomaly. Consequently, interfaces with large $\theta_\mathrm{s}$ are required for a clear observation of bound states. Also, the resonances broaden into a subgap continuum at highly transparent interfaces, so that low-transparency tunnel junctions are required to clearly observe them. Recent theoretical predictions show that $\theta_\mathrm{s}$ can be of the order of $\pi$ in superconductor-ferromagnet junctions with ultra-thin tunnel barriers \cite{grein2010}, and this has motivated us to fabricate this type of junctions.

Samples with multiple junctions were fabricated by standard e-beam lithography and a sequence of oblique-angle (shadow) evaporation techniques. The junctions consist of a thin ($\approx~12-15~\mathrm{nm}$) layer of aluminum, which was oxidized {\it in situ} to form a tunnel barrier, and an iron counterelectrode. Typical parameters of the aluminum films are: resistivity $\rho\approx 10~\mathrm{\mu\Omega cm}$, critical temperature $T_\mathrm{c}\approx 1.5~\mathrm{K}$ and critical field $B_\mathrm{c}\approx 2~\mathrm{T}$ for in-plane magnetic fields. A total of 8 samples were prepared, with 2 to 5 junctions each. Normal-state junction conductances were around $G_\mathrm{N}\approx 1~\mathrm{mS}$, with a typical junction area of $150\times 150~\mathrm{nm}^2$, as shown in Fig.~\ref{fig_scheme}(a). This corresponds to resistance-area products as low as $30~\Omega\mathrm{\mu m}^2$, and an average transmission probability of $\langle t \rangle\approx 5\times 10^{-5}$. In Fig.~\ref{fig_scheme}(b), the structure of the tunnel barrier can be resolved in a high-resolution transmission electron microscopy (HRTEM) image of a reference structure fabricated by the same methods. As can be seen, the oxide barrier is amorphous, and varies in thickness between 1 and 2 nm. Assuming an effective tunnel barrier height of about $1~\mathrm{eV}$, the tunneling probability for electrons through a barrier of 1~nm thickness is about $10^{-5}$, consistent with the observed resistance-area product. The differential conductance of these junctions was measured at low temperatures using a low-frequency ac method with an excitation of $\approx 5~\mathrm{\mu V}$ at $138~\mathrm{Hz}$. Details of the experimental setup have been described elsewhere \cite{brauer2010,huebler2010}.

\begin{figure}
\includegraphics[width=\columnwidth]{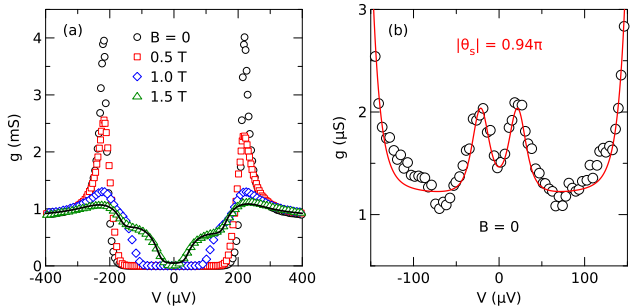}
\caption{\label{fig_dIdV}(Color online)
Differential conductance spectra. (a) Differential conductance of one junction as a function of applied magnetic field $B$ at lowest temperature. The solid line is a fit to (\ref{equ_gtunnel}). (b) subgap differential conductance of the same contact at $B=0$ (symbols), together with a fit according to (\ref{equ_g}) (line).}
\end{figure}

Figure \ref{fig_dIdV}(a) shows the differential conductance of one junction at lowest temperature ($T=50~\mathrm{mK}$) for different magnetic fields applied in-plane parallel to the iron strip. At zero field, the data reveal the density of states of the superconductor with a well-resolved energy gap $\Delta\approx 200~\mathrm{\mu eV}$. With increasing magnetic field, the gap features broaden, and the Zeeman splitting of the density of states for spin up and down is observed \cite{tedrow1971}. Figure \ref{fig_dIdV}(b) shows the subgap differential conductance of the same junction measured at $B=0$ (symbols) on an enlarged scale. Two well-resolved peaks of equal height are seen at $V=\pm 23~\mathrm{\mu V}$ (note the expansion of the vertical scale by a factor of 1500). These peaks exhibit two of the salient features of resonant Andreev reflection via bound states: First, their position is symmetric with respect to the chemical potential of the superconductor. Second, the peak height is symmetric with respect to bias, as expected for resonant Andreev reflection. 

To fit our data, we assume three contributions to the conductance
\begin{equation}
g(V)=g_\mathrm{T}(V)+g_\mathrm{ABS}(V)+g_0. \label{equ_g}
\end{equation}
Here $g_\mathrm{T}(V)$ describes the dominating contribution of $\approx 1~\mathrm{mS}$ for $eV>\Delta$ due to ordinary quasiparticle tunneling, $g_\mathrm{ABS}(V)$ represents the much smaller subgap contributions due to an Andreev bound state, and $g_0\approx1~\mathrm{\mu S}$ accounts for additional subgap leakage through the thin oxide barrier. 

For $g_\mathrm{T}(V)$ we have used the standard model of spin-polarized quasiparticle tunneling  in high magnetic fields \cite{skalski1964,maki1964a,meservey1975}
\begin{equation}
g_\mathrm{T}(V)=\frac{G_\mathrm{N}}{2}\sum_{\pm}(1\mp P)\int n_\pm(\epsilon)f'd\epsilon, \label{equ_gtunnel}
\end{equation}
where $P$ is the spin polarization of the tunnel conductance, $f'=-\partial f(\epsilon+eV)/\partial eV$ is the derivative of the Fermi function, $e$ is the elementary charge, and the normalized density of states per spin in the superconductor is given by
\begin{equation}
n_{\pm}(\epsilon)=\mathrm{Re}\left(\frac{u_\pm}{\sqrt{u_\pm^2-1}}\right).\label{equ_n_pm}
\end{equation}
The complex quantities $u_\pm$ have to be determined from the implicit equation
\begin{equation}
\frac{\epsilon\mp\mu_\mathrm{B}B}{\Delta}=u_\pm-\frac{\Gamma}{\Delta}\frac{u_\pm}{\sqrt{1-u_\pm^2}}
+b_\mathrm{so}\frac{u_\pm-u_\mp}{\sqrt{1-u_\mp^2}},\label{equ_u}
\end{equation}
where $\mu_\mathrm{B}$ is the Bohr magneton, $\Gamma$ is the magnetic pair-breaking parameter and $b_\mathrm{so}=\hbar/3\tau_\mathrm{so}\Delta$ measures the spin-orbit scattering strength. 

Since the observed subgap peaks were always much smaller than the conductance quantum, we have assumed a single conductance channel for $g_\mathrm{ABS}(V)$. The subgap conductance due to an Andreev bound state is then given by \cite{zhao2004}
\begin{equation}
g_\mathrm{ABS}(V)=\frac{e^2}{h}\int\sum_{\pm}\frac{\tau^2}{1+\rho^2-2\rho\cos\left(2\delta\pm\theta_\mathrm{s}\right)}f' d\epsilon, \label{equ_gabs}
\end{equation}
where $\cos\delta=\epsilon/\Delta$, $\tau^2=t_+t_-$, $\rho^2=r_+r_-$, and $t_\pm$ and $r_\pm=1-t_\pm$ are the spin-dependent transmission and reflection probabilities of the interface. 

For fitting the data, we first used only the dominating contribution $g_\mathrm{T}(V)$ to obtain the pair potential $\Delta$, the magnetic pair-breaking parameter $\Gamma$, the normal-state conductance $G_\mathrm{N}$, and the degree of spin polarisation $P$, following the same procedure as described in Ref. \cite{huebler2010}. To obtain a good fit of the onset of the gap features, we also had to adjust the temperature ($T\approx50-100~\mathrm{mK}$), and add a life-time broadening parameter $\Gamma_\mathrm{lt}$ by replacing $\epsilon$ by $\epsilon+i\Gamma_\mathrm{lt}$ in (\ref{equ_u}) in order to optimize the fit. For the spin polarization of the tunnel conductance we obtain $P\approx10-15\%$, typical for junctions with ultra-thin amorphous aluminum oxide barriers \cite{muenzenberg2004}. The same $P$ was also extracted from non-local spin-valve experiments performed at $T=4.2~\mathrm{K}$. From these experiments, we also obtain the coercive field of our Fe wires of about 50~mT.

After fixing $g_\mathrm{T}(V)$, we then added $g_\mathrm{ABS}(V)$ and $g_0$ to fit the subgap peaks. We obtained typical values $t_+\approx 0.1$ and $t_-\approx 10^{-3}$ from our fits for all the observed bound states (note that $t_+$ and $t_-$ are interchangeable in (\ref{equ_gabs}), and we have arbitrarily chosen $t_+>t_-$). The product $t_+ t_-$ essentially fixes the spectral weight, while the larger of the two fixes the width of the conductance peaks. Whether $t_+\gg t_-$ signifies a locally enhanced spin polarization or simply an additional life-time broadening of the resonances can not be determined from the fits. The peak position, and consequently $|\theta_\mathrm{s}|$, varied strongly from junction to junction (see below). The result of such a fit, with $|\theta_\mathrm{s}|=0.94\pi$, is shown in Fig.~\ref{fig_dIdV}(b). Upon increasing temperature, the peaks simply broaden further, until they disappear in the smearing of the gap features at about 100-150~mK (not shown).

\begin{figure}
\includegraphics[width=\columnwidth]{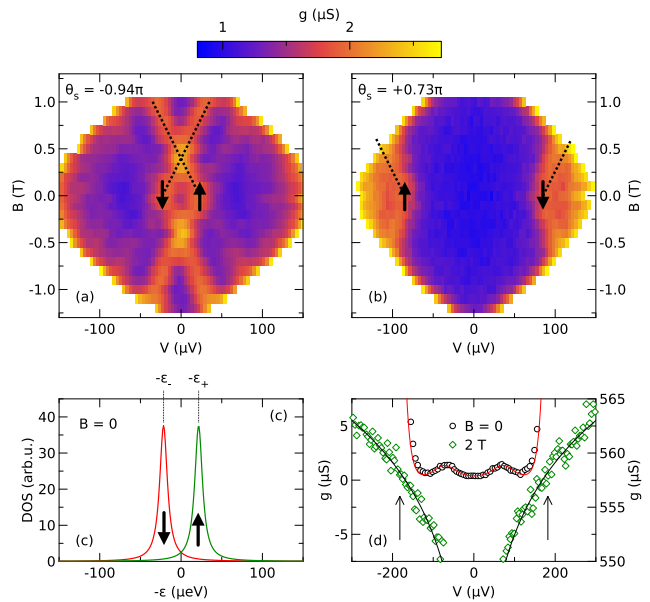}
\caption{\label{fig_Zeeman}(Color online)
Density of states and Zeeman shift of the bound states. 
(a) and (b) Differential subgap conductance of two contacts of the same sample as a function of applied magnetic field $B$. The dotted lines indicate the Zeeman shift of the bound states (shown for positive fields only), arrows indicate spin. 
(c) Spin-resolved density of states due to the bound state shown in panel (a), plotted for the parameters obtained from fits to (\ref{equ_gabs}), and with the sign of the spin-mixing angle inferred from the Zeeman shift (see text).
(d) Data for a third contact at $B=0$ (circles, left conductance scale) and in the normal state at $B=2~\mathrm{T}$ (diamonds, right conductance scale). Both scales span $15~\mathrm{\mu S}$. The lines are fits to (\ref{equ_g}) and the standard model of dynamic Coulomb blockade \cite{devoret1990}, respectively (see text).}
\end{figure}

The density of states induced by Andreev bound states is spin polarized, i.e., the two conductance peaks corresponds to opposite spin. This expectation can be checked by observing the evolution of the peaks with magnetic field, as shown in Fig.~\ref{fig_Zeeman}(a). The peak features can be traced up to $B\approx1~\mathrm{T}$, where they merge with the onset of the gap features given by $g_\mathrm{T}(V)$. The magnetic-field dependence of the pair potential $\Delta$ due to orbital pair breaking obtained from the fits to $g_\mathrm{T}(V)$ is negligible in this field range, so that the shift of the subgap peaks is entirely due to the Zeeman effect. As can be seen, the peaks shift linearly into opposite directions with a slope of $\pm\mu_\mathrm{B}B$, as indicated by the dotted lines. From the direction of the Zeeman shift, we can unambiguously assign the spin of the conductance features, as indicated by arrows. We can then use the spin assignment to infer the sign of the phase shift $\theta_\mathrm{s}=-0.94\pi$. To illustrate this, we have also plotted the spin-dependent density of states \cite{zhao2004} in Fig.~\ref{fig_Zeeman}(c) for the same parameters as used for the fit shown in Fig.~\ref{fig_dIdV}(b). In Fig.~\ref{fig_Zeeman}(b), we display the data for a second contact for comparison. As can be seen, the energy of the bound state differs, and also the Zeeman shift is in the opposite direction. Consequently, we can assign a positive $\theta_\mathrm{s}=+0.73\pi$. Since the coercive field of the iron strips of about 50~mT is smaller than the magnetic-field increments in Fig.~\ref{fig_Zeeman}, the magnetization and magnetic field are always parallel for both $B>0$ and $B<0$. Consequently, the Zeeman shift seen for $B>0$ is mirrored for $B<0$, since spin up and down simply exchange their roles in the upper and lower half-plane. This is consistent with the view that the observed peaks are induced by the leakage of the exchange field of the ferromagnet into the superconductor. We also note that the observed magnetic-field dependence of the peaks excludes the possibility that they are minigap features due to the proximity effect, since these should quickly disappear at higher fields. To check whether the peaks might be due to some inelastic processes unrelated to superconductivity, Fig.~\ref{fig_Zeeman}(d) compares the peaks seen in a third contact in the superconducting state at $B=0$ to the normal state data at $B=2~\mathrm{T}$ just above the critical field. In the normal state, a Coulomb dip is observed, which can be fit with the standard model \cite{devoret1990}. The arrows indicate the extrapolation of the peak positions from $B=0$ to $B=2~\mathrm{T}$, using the Zeeman shift observed in the superconducting state. No peaks are seen at these positions in the normal state within the resolution.

\begin{figure}
\includegraphics[width=\columnwidth]{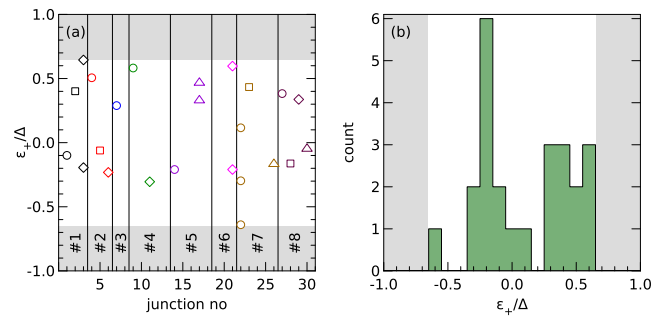}
\caption{\label{fig_theta}(Color online)
(a) Normalized bound-state energy $\epsilon_+/\Delta$ for all measured junctions. Sample numbers are given at the bottom.
(b) Histogram of the data from panel (a), bin size 0.1.}
\end{figure}

A total of 30 junctions were measured, with results summarized in Fig.~\ref{fig_theta}. In Fig.~\ref{fig_theta}(a), we show the normalized energy $\epsilon_+/\Delta$ of the spin-up feature, as determined from the fits and the Zeeman shift. In 19 of the junctions usually a single bound state was observed, and in some cases, there were two or three states in a single junction (e.g., contacts 21 and 22). 11 junctions did not show any subgap features, which means that there is either no bound state, or its energy is too close to $\Delta$ to be distinguished from the onset of the gap features. The energy range where observation is impeded is greyed out in Fig.~\ref{fig_theta}(a). The corresponding histogram of bound-state energies is shown in Fig.~\ref{fig_theta}(b). As can be seen, there is a broad distribution, with maybe a slight preference for values around $\epsilon_+/\Delta\approx -0.2$ and $0.4$, corresponding to $\theta_\mathrm{s}\approx -0.9\pi$ and $+0.75\pi$.

From these findings the following picture emerges: The simultaneous presence of strong gap features and small overall subgap conductance indicates that over most of the area of a given junction both $\theta_\mathrm{s}$ and the transmission probability are small. Only for a small part of the junction, where the oxide barrier is thinnest, $\theta_\mathrm{s}$ is large, and the transmission probability is sufficiently large ($t \gtrsim 10^{-3}$) to sustain Andreev transport. This is corroborated by the thickness variation of the oxide barrier seen in the HRTEM image in Fig.~\ref{fig_scheme}(b). Recent theoretical predictions suggest that $\theta_\mathrm{s}$ can be large for an ultra-thin tunnel barrier with a smooth potential profile \cite{grein2010}. The broad distribution of $\theta_\mathrm{s}$ reflects the extreme sensitivity of spin mixing to interface properties. 

A possible alternative scenario for the appearance of subgap Andreev resonances would be the presence of discrete impurity states in the disordered oxide barrier, effectively acting as quantum dots (see, e.g., \cite{domanski2008,koerting2010}, and references therein). This might easily explain why we see discrete resonances despite the fact that we have large-area disordered contacts. However, a couple of observations are incompatible with this view. First, such features were not observed in previous experiments on non-magnetic junctions \cite{brauer2010,huebler2010} fabricated in the same way and with similar resistance-area product, where the same type and density of impurity states should be present. Second, in this case a Zeeman splitting rather than a simple shift of the resonances would be expected in the magnetic field \cite{domanski2008}. Third, both the energy and the width of the resonances would depend on the coupling strength to the superconducting electrode, with broader features at lower energy \cite{koerting2010}. In contrast, we did not observe a systematic dependence of the peak width on energy within the scatter of the data. And finally, magnetic impurity states at the surface of a superconductor usually exhibit an asymmetry in peak height \cite{franke2011}, which we did not observe. Since the features are very small, one might also imagine a weak remnant of inelastic single-particle tunneling via magnetic impurity states \cite{shen1968,*wolf1970}. This would however be subject to the same thermal suppression as the overall subgap conductance, and should therefore be clearly visible in the normal-state conductance at high magnetic fields, where no features were observed at the expected positions in the experiment.

In conclusion, we have shown evidence for subgap Andreev bound states at spin-active interfaces between superconductors and ferromagnets, and directly determined the spin-dependent interface phase shift $\theta_\mathrm{s}$, thus providing microscopic insight into the generation of triplet superconductivity at spin-active interfaces. We find that $\theta_\mathrm{s}$ can be surprisingly large, of the order of $\pi$, in agreement with recent theoretical predictions for ultra-thin tunnel barriers.

We would like to thank R. Grein, G. Metalidis, M. Eschrig, W. Belzig and M. Aprili for valuable discussions, and M. Smetanin for x-ray characterization of our samples. This work was partially supported by the Deutsche Forschungsgemeinschaft.

% Create the reference section using BibTeX:
\bibliography{../../../lit.bib}

%merlin.mbs 2010-03-15 4.21a (PWD, AO, DPC)
%Control: key (0)
%Control: author (8) initials jnrlst
%Control: editor formatted (1) identically to author
%Control: production of article title (-1) disabled
%Control: page (0) single
%Control: year (1) truncated
%Control: production of eprint (0) enabled
\begin{thebibliography}{37}%
\makeatletter
\providecommand \@ifxundefined [1]{%
 \@ifx{#1\undefined}
}%
\providecommand \@ifnum [1]{%
 \ifnum #1\expandafter \@firstoftwo
 \else \expandafter \@secondoftwo
 \fi
}%
\providecommand \@ifx [1]{%
 \ifx #1\expandafter \@firstoftwo
 \else \expandafter \@secondoftwo
 \fi
}%
\providecommand \natexlab [1]{#1}%
\providecommand \enquote  [1]{``#1''}%
\providecommand \bibnamefont  [1]{#1}%
\providecommand \bibfnamefont [1]{#1}%
\providecommand \citenamefont [1]{#1}%
\providecommand \href@noop [0]{\@secondoftwo}%
\providecommand \href [0]{\begingroup \@sanitize@url \@href}%
\providecommand \@href[1]{\@@startlink{#1}\@@href}%
\providecommand \@@href[1]{\endgroup#1\@@endlink}%
\providecommand \@sanitize@url [0]{\catcode `\\12\catcode `\$12\catcode
  `\&12\catcode `\#12\catcode `\^12\catcode `\_12\catcode `\%12\relax}%
\providecommand \@@startlink[1]{}%
\providecommand \@@endlink[0]{}%
\providecommand \url  [0]{\begingroup\@sanitize@url \@url }%
\providecommand \@url [1]{\endgroup\@href {#1}{\urlprefix }}%
\providecommand \urlprefix  [0]{URL }%
\providecommand \Eprint [0]{\href }%
\@ifxundefined \urlstyle {%
  \providecommand \doi  [0]{\begingroup \@sanitize@url \@doi}%
  \providecommand \@doi [1]{\endgroup \@@startlink {\doibase
  #1}doi:\discretionary {}{}{}#1\@@endlink }%
}{%
  \providecommand \doi  [0]{doi:\discretionary{}{}{}\begingroup
  \urlstyle{rm}\Url }%
}%
\providecommand \doibase [0]{http://dx.doi.org/}%
\providecommand \Doi [0]{\begingroup \@sanitize@url \@Doi }%
\providecommand \@Doi  [1]{\endgroup\@@startlink{\doibase#1}\@@Doi}%
\providecommand \@@Doi [1]{#1\@@endlink}%
\providecommand \selectlanguage [0]{\@gobble}%
\providecommand \bibinfo  [0]{\@secondoftwo}%
\providecommand \bibfield  [0]{\@secondoftwo}%
\providecommand \translation [1]{[#1]}%
\providecommand \BibitemOpen [0]{}%
\providecommand \bibitemStop [0]{}%
\providecommand \bibitemNoStop [0]{.\EOS\space}%
\providecommand \EOS [0]{\spacefactor3000\relax}%
\providecommand \BibitemShut  [1]{\csname bibitem#1\endcsname}%
%</preamble>
\bibitem [{\citenamefont {Keizer}\ \emph {et~al.}(2006)\citenamefont {Keizer},
  \citenamefont {Goennenwein}, \citenamefont {Klapwijk}, \citenamefont {Miao},
  \citenamefont {Xiao},\ and\ \citenamefont {Gupta}}]{keizer2006}%
  \BibitemOpen
  \bibfield  {author} {\bibinfo {author} {\bibfnamefont {R.~S.}\ \bibnamefont
  {Keizer}}, \bibinfo {author} {\bibfnamefont {S.~T.~B.}\ \bibnamefont
  {Goennenwein}}, \bibinfo {author} {\bibfnamefont {T.~M.}\ \bibnamefont
  {Klapwijk}}, \bibinfo {author} {\bibfnamefont {G.}~\bibnamefont {Miao}},
  \bibinfo {author} {\bibfnamefont {G.}~\bibnamefont {Xiao}}, \ and\ \bibinfo
  {author} {\bibfnamefont {A.}~\bibnamefont {Gupta}},\ }\href
  {http://www.nature.com/nature/journal/v439/n7078/abs/nature04499.html}
  {\bibinfo  {journal} {Nature} \textbf {\bibinfo {volume} {439}},\ \bibinfo
  {pages} {825} (\bibinfo {year} {2006})}\BibitemShut {NoStop}%
\bibitem [{\citenamefont {Bergeret}\ \emph {et~al.}(2001)\citenamefont
  {Bergeret}, \citenamefont {Volkov},\ and\ \citenamefont
  {Efetov}}]{bergeret2001}%
  \BibitemOpen
  \bibfield  {author} {\bibinfo {author} {\bibfnamefont {F.~S.}\ \bibnamefont
  {Bergeret}}, \bibinfo {author} {\bibfnamefont {A.~F.}\ \bibnamefont
  {Volkov}}, \ and\ \bibinfo {author} {\bibfnamefont {K.~B.}\ \bibnamefont
  {Efetov}},\ }\href {http://link.aps.org/abstract/PRL/v86/p4096} {\bibinfo
  {journal} {Phys. Rev. Lett.} \textbf {\bibinfo {volume} {86}},\ \bibinfo
  {pages} {4096} (\bibinfo {year} {2001})}\BibitemShut {NoStop}%
\bibitem [{\citenamefont {Bergeret}\ \emph
  {et~al.}(2005){\natexlab{a}}\citenamefont {Bergeret}, \citenamefont
  {Volkov},\ and\ \citenamefont {Efetov}}]{bergeret2005}%
  \BibitemOpen
  \bibfield  {author} {\bibinfo {author} {\bibfnamefont {F.~S.}\ \bibnamefont
  {Bergeret}}, \bibinfo {author} {\bibfnamefont {A.~F.}\ \bibnamefont
  {Volkov}}, \ and\ \bibinfo {author} {\bibfnamefont {K.~B.}\ \bibnamefont
  {Efetov}},\ }\href {http://dx.doi.org/10.1103/RevModPhys.77.1321} {\bibinfo
  {journal} {Rev. Mod. Phys.} \textbf {\bibinfo {volume} {77}},\ \bibinfo
  {pages} {1321} (\bibinfo {year} {2005}{\natexlab{a}})}\BibitemShut {NoStop}%
\bibitem [{\citenamefont {Buzdin}(2005)}]{buzdin2005}%
  \BibitemOpen
  \bibfield  {author} {\bibinfo {author} {\bibfnamefont {A.~I.}\ \bibnamefont
  {Buzdin}},\ }\href {http://dx.doi.org/10.1103/RevModPhys.77.935} {\bibinfo
  {journal} {Rev. Mod. Phys.} \textbf {\bibinfo {volume} {77}},\ \bibinfo
  {pages} {935} (\bibinfo {year} {2005})}\BibitemShut {NoStop}%
\bibitem [{\citenamefont {Millis}\ \emph {et~al.}(1988)\citenamefont {Millis},
  \citenamefont {Rainer},\ and\ \citenamefont {Sauls}}]{millis1988}%
  \BibitemOpen
  \bibfield  {author} {\bibinfo {author} {\bibfnamefont {A.}~\bibnamefont
  {Millis}}, \bibinfo {author} {\bibfnamefont {D.}~\bibnamefont {Rainer}}, \
  and\ \bibinfo {author} {\bibfnamefont {J.~A.}\ \bibnamefont {Sauls}},\ }\href
  {http://link.aps.org/abstract/PRB/v38/p4504} {\bibinfo  {journal} {Phys. Rev.
  B} \textbf {\bibinfo {volume} {38}},\ \bibinfo {pages} {4504} (\bibinfo
  {year} {1988})}\BibitemShut {NoStop}%
\bibitem [{\citenamefont {Eschrig}\ and\ \citenamefont
  {L\"ofwander}(2008)}]{eschrig2008}%
  \BibitemOpen
  \bibfield  {author} {\bibinfo {author} {\bibfnamefont {M.}~\bibnamefont
  {Eschrig}}\ and\ \bibinfo {author} {\bibfnamefont {T.}~\bibnamefont
  {L\"ofwander}},\ }\href
  {http://www.nature.com/nphys/journal/v4/n2/abs/nphys831.html} {\bibinfo
  {journal} {Nature Phys.} \textbf {\bibinfo {volume} {4}},\ \bibinfo {pages}
  {138} (\bibinfo {year} {2008})}\BibitemShut {NoStop}%
\bibitem [{\citenamefont {Kontos}\ \emph {et~al.}(2002)\citenamefont {Kontos},
  \citenamefont {Aprili}, \citenamefont {Lesueur}, \citenamefont {Genet},
  \citenamefont {Stephanidis},\ and\ \citenamefont {Boursier}}]{kontos2002}%
  \BibitemOpen
  \bibfield  {author} {\bibinfo {author} {\bibfnamefont {T.}~\bibnamefont
  {Kontos}}, \bibinfo {author} {\bibfnamefont {M.}~\bibnamefont {Aprili}},
  \bibinfo {author} {\bibfnamefont {J.}~\bibnamefont {Lesueur}}, \bibinfo
  {author} {\bibfnamefont {F.}~\bibnamefont {Genet}}, \bibinfo {author}
  {\bibfnamefont {B.}~\bibnamefont {Stephanidis}}, \ and\ \bibinfo {author}
  {\bibfnamefont {R.}~\bibnamefont {Boursier}},\ }\href
  {http://link.aps.org/abstract/PRL/v89/e137007} {\bibinfo  {journal} {Phys.
  Rev. Lett.} \textbf {\bibinfo {volume} {89}},\ \bibinfo {pages} {137007}
  (\bibinfo {year} {2002})}\BibitemShut {NoStop}%
\bibitem [{\citenamefont {Cottet}\ and\ \citenamefont
  {Belzig}(2005)}]{cottet2005}%
  \BibitemOpen
  \bibfield  {author} {\bibinfo {author} {\bibfnamefont {A.}~\bibnamefont
  {Cottet}}\ and\ \bibinfo {author} {\bibfnamefont {W.}~\bibnamefont
  {Belzig}},\ }\href {http://link.aps.org/abstract/PRB/v72/e180503} {\bibinfo
  {journal} {Phys. Rev. B} \textbf {\bibinfo {volume} {72}},\ \bibinfo {pages}
  {180503} (\bibinfo {year} {2005})}\BibitemShut {NoStop}%
\bibitem [{\citenamefont {Khaire}\ \emph {et~al.}(2010)\citenamefont {Khaire},
  \citenamefont {Khasawneh}, \citenamefont {Pratt},\ and\ \citenamefont
  {Birge}}]{khaire2010}%
  \BibitemOpen
  \bibfield  {author} {\bibinfo {author} {\bibfnamefont {T.~S.}\ \bibnamefont
  {Khaire}}, \bibinfo {author} {\bibfnamefont {M.~A.}\ \bibnamefont
  {Khasawneh}}, \bibinfo {author} {\bibfnamefont {W.~P.}\ \bibnamefont
  {Pratt}}, \ and\ \bibinfo {author} {\bibfnamefont {N.~O.}\ \bibnamefont
  {Birge}},\ }\href {http://link.aps.org/doi/10.1103/PhysRevLett.104.137002}
  {\bibinfo  {journal} {Phys. Rev. Lett.} \textbf {\bibinfo {volume} {104}},\
  \bibinfo {pages} {137002} (\bibinfo {year} {2010})}\BibitemShut {NoStop}%
\bibitem [{\citenamefont {Robinson}\ \emph {et~al.}(2010)\citenamefont
  {Robinson}, \citenamefont {Witt},\ and\ \citenamefont
  {Blamire}}]{robinson2010}%
  \BibitemOpen
  \bibfield  {author} {\bibinfo {author} {\bibfnamefont {J.~W.~A.}\
  \bibnamefont {Robinson}}, \bibinfo {author} {\bibfnamefont {J.~D.~S.}\
  \bibnamefont {Witt}}, \ and\ \bibinfo {author} {\bibfnamefont {M.~G.}\
  \bibnamefont {Blamire}},\ }\href
  {http://www.sciencemag.org/cgi/content/abstract/329/5987/59} {\bibinfo
  {journal} {Science} \textbf {\bibinfo {volume} {329}},\ \bibinfo {pages} {59}
  (\bibinfo {year} {2010})}\BibitemShut {NoStop}%
\bibitem [{\citenamefont {Anwar}\ \emph {et~al.}(2010)\citenamefont {Anwar},
  \citenamefont {Czeschka}, \citenamefont {Hesselberth}, \citenamefont
  {Porcu},\ and\ \citenamefont {Aarts}}]{anwar2010}%
  \BibitemOpen
  \bibfield  {author} {\bibinfo {author} {\bibfnamefont {M.~S.}\ \bibnamefont
  {Anwar}}, \bibinfo {author} {\bibfnamefont {F.}~\bibnamefont {Czeschka}},
  \bibinfo {author} {\bibfnamefont {M.}~\bibnamefont {Hesselberth}}, \bibinfo
  {author} {\bibfnamefont {M.}~\bibnamefont {Porcu}}, \ and\ \bibinfo {author}
  {\bibfnamefont {J.}~\bibnamefont {Aarts}},\ }\href
  {http://link.aps.org/doi/10.1103/PhysRevB.82.100501} {\bibinfo  {journal}
  {Phys. Rev. B} \textbf {\bibinfo {volume} {82}},\ \bibinfo {pages} {100501}
  (\bibinfo {year} {2010})}\BibitemShut {NoStop}%
\bibitem [{\citenamefont {L\"ofwander}\ \emph {et~al.}(2010)\citenamefont
  {L\"ofwander}, \citenamefont {Grein},\ and\ \citenamefont
  {Eschrig}}]{loefwander2010}%
  \BibitemOpen
  \bibfield  {author} {\bibinfo {author} {\bibfnamefont {T.}~\bibnamefont
  {L\"ofwander}}, \bibinfo {author} {\bibfnamefont {R.}~\bibnamefont {Grein}},
  \ and\ \bibinfo {author} {\bibfnamefont {M.}~\bibnamefont {Eschrig}},\ }\href
  {http://link.aps.org/doi/10.1103/PhysRevLett.105.207001} {\bibinfo  {journal}
  {Phys. Rev. Lett.} \textbf {\bibinfo {volume} {105}},\ \bibinfo {pages}
  {207001} (\bibinfo {year} {2010})}\BibitemShut {NoStop}%
\bibitem [{\citenamefont {Usman}\ \emph {et~al.}(2011)\citenamefont {Usman},
  \citenamefont {Yates}, \citenamefont {Moore}, \citenamefont {Morrison},
  \citenamefont {Pecharsky}, \citenamefont {Gschneidner}, \citenamefont
  {Verhagen}, \citenamefont {Aarts}, \citenamefont {Zverev}, \citenamefont
  {Robinson}, \citenamefont {Witt}, \citenamefont {Blamire},\ and\
  \citenamefont {Cohen}}]{usman2011}%
  \BibitemOpen
  \bibfield  {author} {\bibinfo {author} {\bibfnamefont {I.~T.~M.}\
  \bibnamefont {Usman}}, \bibinfo {author} {\bibfnamefont {K.~A.}\ \bibnamefont
  {Yates}}, \bibinfo {author} {\bibfnamefont {J.~D.}\ \bibnamefont {Moore}},
  \bibinfo {author} {\bibfnamefont {K.}~\bibnamefont {Morrison}}, \bibinfo
  {author} {\bibfnamefont {V.~K.}\ \bibnamefont {Pecharsky}}, \bibinfo {author}
  {\bibfnamefont {K.~A.}\ \bibnamefont {Gschneidner}}, \bibinfo {author}
  {\bibfnamefont {T.}~\bibnamefont {Verhagen}}, \bibinfo {author}
  {\bibfnamefont {J.}~\bibnamefont {Aarts}}, \bibinfo {author} {\bibfnamefont
  {V.~I.}\ \bibnamefont {Zverev}}, \bibinfo {author} {\bibfnamefont {J.~W.~A.}\
  \bibnamefont {Robinson}}, \bibinfo {author} {\bibfnamefont {J.~D.~S.}\
  \bibnamefont {Witt}}, \bibinfo {author} {\bibfnamefont {M.~G.}\ \bibnamefont
  {Blamire}}, \ and\ \bibinfo {author} {\bibfnamefont {L.~F.}\ \bibnamefont
  {Cohen}},\ }\href {http://link.aps.org/doi/10.1103/PhysRevB.83.144518}
  {\bibinfo  {journal} {Phys. Rev. B} \textbf {\bibinfo {volume} {83}},\
  \bibinfo {pages} {144518} (\bibinfo {year} {2011})}\BibitemShut {NoStop}%
\bibitem [{\citenamefont {Deutscher}(2005)}]{deutscher2005}%
  \BibitemOpen
  \bibfield  {author} {\bibinfo {author} {\bibfnamefont {G.}~\bibnamefont
  {Deutscher}},\ }\href {http://link.aps.org/abstract/RMP/v77/p109} {\bibinfo
  {journal} {Rev. Mod. Phys.} \textbf {\bibinfo {volume} {77}},\ \bibinfo
  {pages} {109} (\bibinfo {year} {2005})}\BibitemShut {NoStop}%
\bibitem [{\citenamefont {Tanaka}\ and\ \citenamefont
  {Kashiwaya}(1995)}]{tanaka1995}%
  \BibitemOpen
  \bibfield  {author} {\bibinfo {author} {\bibfnamefont {Y.}~\bibnamefont
  {Tanaka}}\ and\ \bibinfo {author} {\bibfnamefont {S.}~\bibnamefont
  {Kashiwaya}},\ }\href {http://link.aps.org/doi/10.1103/PhysRevLett.74.3451}
  {\bibinfo  {journal} {Phys. Rev. Lett.} \textbf {\bibinfo {volume} {74}},\
  \bibinfo {pages} {3451} (\bibinfo {year} {1995})}\BibitemShut {NoStop}%
\bibitem [{\citenamefont {Kashiwaya}\ \emph {et~al.}(1995)\citenamefont
  {Kashiwaya}, \citenamefont {Tanaka}, \citenamefont {Koyanagi}, \citenamefont
  {Takashima},\ and\ \citenamefont {Kajimura}}]{kashiwaya1995}%
  \BibitemOpen
  \bibfield  {author} {\bibinfo {author} {\bibfnamefont {S.}~\bibnamefont
  {Kashiwaya}}, \bibinfo {author} {\bibfnamefont {Y.}~\bibnamefont {Tanaka}},
  \bibinfo {author} {\bibfnamefont {M.}~\bibnamefont {Koyanagi}}, \bibinfo
  {author} {\bibfnamefont {H.}~\bibnamefont {Takashima}}, \ and\ \bibinfo
  {author} {\bibfnamefont {K.}~\bibnamefont {Kajimura}},\ }\href
  {http://link.aps.org/abstract/PRB/v51/p1350} {\bibinfo  {journal} {Phys. Rev.
  B} \textbf {\bibinfo {volume} {51}},\ \bibinfo {pages} {1350} (\bibinfo
  {year} {1995})}\BibitemShut {NoStop}%
\bibitem [{\citenamefont {Yamashiro}\ \emph {et~al.}(1997)\citenamefont
  {Yamashiro}, \citenamefont {Tanaka},\ and\ \citenamefont
  {Kashiwaya}}]{yamashiro1997}%
  \BibitemOpen
  \bibfield  {author} {\bibinfo {author} {\bibfnamefont {M.}~\bibnamefont
  {Yamashiro}}, \bibinfo {author} {\bibfnamefont {Y.}~\bibnamefont {Tanaka}}, \
  and\ \bibinfo {author} {\bibfnamefont {S.}~\bibnamefont {Kashiwaya}},\ }\href
  {http://link.aps.org/doi/10.1103/PhysRevB.56.7847} {\bibinfo  {journal}
  {Phys. Rev. B} \textbf {\bibinfo {volume} {56}},\ \bibinfo {pages} {7847}
  (\bibinfo {year} {1997})}\BibitemShut {NoStop}%
\bibitem [{\citenamefont {Laube}\ \emph {et~al.}(2000)\citenamefont {Laube},
  \citenamefont {Goll}, \citenamefont {L\"ohneysen}, \citenamefont
  {Fogelstr\"om},\ and\ \citenamefont {Lichtenberg}}]{laube2000}%
  \BibitemOpen
  \bibfield  {author} {\bibinfo {author} {\bibfnamefont {F.}~\bibnamefont
  {Laube}}, \bibinfo {author} {\bibfnamefont {G.}~\bibnamefont {Goll}},
  \bibinfo {author} {\bibfnamefont {H.~v.}\ \bibnamefont {L\"ohneysen}},
  \bibinfo {author} {\bibfnamefont {M.}~\bibnamefont {Fogelstr\"om}}, \ and\
  \bibinfo {author} {\bibfnamefont {F.}~\bibnamefont {Lichtenberg}},\ }\href
  {http://link.aps.org/doi/10.1103/PhysRevLett.84.1595} {\bibinfo  {journal}
  {Phys. Rev. Lett.} \textbf {\bibinfo {volume} {84}},\ \bibinfo {pages} {1595}
  (\bibinfo {year} {2000})}\BibitemShut {NoStop}%
\bibitem [{\citenamefont {Zhao}\ \emph {et~al.}(2004)\citenamefont {Zhao},
  \citenamefont {L\"ofwander},\ and\ \citenamefont {Sauls}}]{zhao2004}%
  \BibitemOpen
  \bibfield  {author} {\bibinfo {author} {\bibfnamefont {E.}~\bibnamefont
  {Zhao}}, \bibinfo {author} {\bibfnamefont {T.}~\bibnamefont {L\"ofwander}}, \
  and\ \bibinfo {author} {\bibfnamefont {J.~A.}\ \bibnamefont {Sauls}},\ }\href
  {http://link.aps.org/abstract/PRB/v70/e134510} {\bibinfo  {journal} {Phys.
  Rev. B} \textbf {\bibinfo {volume} {70}},\ \bibinfo {pages} {134510}
  (\bibinfo {year} {2004})}\BibitemShut {NoStop}%
\bibitem [{\citenamefont {Bergeret}\ \emph
  {et~al.}(2005){\natexlab{b}}\citenamefont {Bergeret}, \citenamefont
  {Levy~Yeyati},\ and\ \citenamefont {Mart\'{\i}n-Rodero}}]{bergeret2005b}%
  \BibitemOpen
  \bibfield  {author} {\bibinfo {author} {\bibfnamefont {F.~S.}\ \bibnamefont
  {Bergeret}}, \bibinfo {author} {\bibfnamefont {A.}~\bibnamefont
  {Levy~Yeyati}}, \ and\ \bibinfo {author} {\bibfnamefont {A.}~\bibnamefont
  {Mart\'{\i}n-Rodero}},\ }\href
  {http://link.aps.org/doi/10.1103/PhysRevB.72.064524} {\bibinfo  {journal}
  {Phys. Rev. B} \textbf {\bibinfo {volume} {72}},\ \bibinfo {pages} {064524}
  (\bibinfo {year} {2005}{\natexlab{b}})}\BibitemShut {NoStop}%
\bibitem [{\citenamefont {Grein}\ \emph {et~al.}(2010)\citenamefont {Grein},
  \citenamefont {L\"ofwander}, \citenamefont {Metalidis},\ and\ \citenamefont
  {Eschrig}}]{grein2010}%
  \BibitemOpen
  \bibfield  {author} {\bibinfo {author} {\bibfnamefont {R.}~\bibnamefont
  {Grein}}, \bibinfo {author} {\bibfnamefont {T.}~\bibnamefont {L\"ofwander}},
  \bibinfo {author} {\bibfnamefont {G.}~\bibnamefont {Metalidis}}, \ and\
  \bibinfo {author} {\bibfnamefont {M.}~\bibnamefont {Eschrig}},\ }\href
  {http://link.aps.org/doi/10.1103/PhysRevB.81.094508} {\bibinfo  {journal}
  {Phys. Rev. B} \textbf {\bibinfo {volume} {81}},\ \bibinfo {pages} {094508}
  (\bibinfo {year} {2010})}\BibitemShut {NoStop}%
\bibitem [{\citenamefont {Kalenkov}\ and\ \citenamefont
  {Zaikin}(2007)}]{kalenkov2007}%
  \BibitemOpen
  \bibfield  {author} {\bibinfo {author} {\bibfnamefont {M.~S.}\ \bibnamefont
  {Kalenkov}}\ and\ \bibinfo {author} {\bibfnamefont {A.~D.}\ \bibnamefont
  {Zaikin}},\ }\href {http://link.aps.org/abstract/PRB/v76/e224506} {\bibinfo
  {journal} {Phys. Rev. B} \textbf {\bibinfo {volume} {76}},\ \bibinfo {pages}
  {224506} (\bibinfo {year} {2007})}\BibitemShut {NoStop}%
\bibitem [{\citenamefont {Metalidis}\ \emph {et~al.}(2010)\citenamefont
  {Metalidis}, \citenamefont {Eschrig}, \citenamefont {Grein},\ and\
  \citenamefont {Sch\"on}}]{metalidis2010}%
  \BibitemOpen
  \bibfield  {author} {\bibinfo {author} {\bibfnamefont {G.}~\bibnamefont
  {Metalidis}}, \bibinfo {author} {\bibfnamefont {M.}~\bibnamefont {Eschrig}},
  \bibinfo {author} {\bibfnamefont {R.}~\bibnamefont {Grein}}, \ and\ \bibinfo
  {author} {\bibfnamefont {G.}~\bibnamefont {Sch\"on}},\ }\href
  {http://link.aps.org/doi/10.1103/PhysRevB.82.180503} {\bibinfo  {journal}
  {Phys. Rev. B} \textbf {\bibinfo {volume} {82}},\ \bibinfo {pages} {180503}
  (\bibinfo {year} {2010})}\BibitemShut {NoStop}%
\bibitem [{\citenamefont {Cottet}\ and\ \citenamefont
  {Belzig}(2008)}]{cottet2008}%
  \BibitemOpen
  \bibfield  {author} {\bibinfo {author} {\bibfnamefont {A.}~\bibnamefont
  {Cottet}}\ and\ \bibinfo {author} {\bibfnamefont {W.}~\bibnamefont
  {Belzig}},\ }\href {http://link.aps.org/abstract/PRB/v77/e064517} {\bibinfo
  {journal} {Phys. Rev. B} \textbf {\bibinfo {volume} {77}},\ \bibinfo {pages}
  {064517} (\bibinfo {year} {2008})}\BibitemShut {NoStop}%
\bibitem [{\citenamefont {Brauer}\ \emph {et~al.}(2010)\citenamefont {Brauer},
  \citenamefont {H\"ubler}, \citenamefont {Smetanin}, \citenamefont
  {Beckmann},\ and\ \citenamefont {v.~L\"ohneysen}}]{brauer2010}%
  \BibitemOpen
  \bibfield  {author} {\bibinfo {author} {\bibfnamefont {J.}~\bibnamefont
  {Brauer}}, \bibinfo {author} {\bibfnamefont {F.}~\bibnamefont {H\"ubler}},
  \bibinfo {author} {\bibfnamefont {M.}~\bibnamefont {Smetanin}}, \bibinfo
  {author} {\bibfnamefont {D.}~\bibnamefont {Beckmann}}, \ and\ \bibinfo
  {author} {\bibfnamefont {H.}~\bibnamefont {v.~L\"ohneysen}},\ }\href
  {http://link.aps.org/doi/10.1103/PhysRevB.81.024515} {\bibinfo  {journal}
  {Phys. Rev. B} \textbf {\bibinfo {volume} {81}},\ \bibinfo {pages} {024515}
  (\bibinfo {year} {2010})}\BibitemShut {NoStop}%
\bibitem [{\citenamefont {H\"ubler}\ \emph {et~al.}(2010)\citenamefont
  {H\"ubler}, \citenamefont {Camirand~Lemyre}, \citenamefont {Beckmann},\ and\
  \citenamefont {v.~L\"ohneysen}}]{huebler2010}%
  \BibitemOpen
  \bibfield  {author} {\bibinfo {author} {\bibfnamefont {F.}~\bibnamefont
  {H\"ubler}}, \bibinfo {author} {\bibfnamefont {J.}~\bibnamefont
  {Camirand~Lemyre}}, \bibinfo {author} {\bibfnamefont {D.}~\bibnamefont
  {Beckmann}}, \ and\ \bibinfo {author} {\bibfnamefont {H.}~\bibnamefont
  {v.~L\"ohneysen}},\ }\href
  {http://link.aps.org/doi/10.1103/PhysRevB.81.184524} {\bibinfo  {journal}
  {Phys. Rev. B} \textbf {\bibinfo {volume} {81}},\ \bibinfo {pages} {184524}
  (\bibinfo {year} {2010})}\BibitemShut {NoStop}%
\bibitem [{\citenamefont {Tedrow}\ and\ \citenamefont
  {Meservey}(1971)}]{tedrow1971}%
  \BibitemOpen
  \bibfield  {author} {\bibinfo {author} {\bibfnamefont {P.~M.}\ \bibnamefont
  {Tedrow}}\ and\ \bibinfo {author} {\bibfnamefont {R.}~\bibnamefont
  {Meservey}},\ }\href {http://link.aps.org/abstract/PRL/v26/p192} {\bibinfo
  {journal} {Phys. Rev. Lett.} \textbf {\bibinfo {volume} {26}},\ \bibinfo
  {pages} {192} (\bibinfo {year} {1971})}\BibitemShut {NoStop}%
\bibitem [{\citenamefont {Skalski}\ \emph {et~al.}(1964)\citenamefont
  {Skalski}, \citenamefont {Betbeder-Matibet},\ and\ \citenamefont
  {Weiss}}]{skalski1964}%
  \BibitemOpen
  \bibfield  {author} {\bibinfo {author} {\bibfnamefont {S.}~\bibnamefont
  {Skalski}}, \bibinfo {author} {\bibfnamefont {O.}~\bibnamefont
  {Betbeder-Matibet}}, \ and\ \bibinfo {author} {\bibfnamefont {P.~R.}\
  \bibnamefont {Weiss}},\ }\href
  {http://link.aps.org/doi/10.1103/PhysRev.136.A1500} {\bibinfo  {journal}
  {Phys. Rev.} \textbf {\bibinfo {volume} {136}},\ \bibinfo {pages} {A1500}
  (\bibinfo {year} {1964})}\BibitemShut {NoStop}%
\bibitem [{\citenamefont {Maki}(1964)}]{maki1964a}%
  \BibitemOpen
  \bibfield  {author} {\bibinfo {author} {\bibfnamefont {K.}~\bibnamefont
  {Maki}},\ }\href {http://ptp.ipap.jp/link?PTP/31/731/} {\bibinfo  {journal}
  {Prog. Theor. Phys.} \textbf {\bibinfo {volume} {31}},\ \bibinfo {pages}
  {731} (\bibinfo {year} {1964})}\BibitemShut {NoStop}%
\bibitem [{\citenamefont {Meservey}\ \emph {et~al.}(1975)\citenamefont
  {Meservey}, \citenamefont {Tedrow},\ and\ \citenamefont
  {Bruno}}]{meservey1975}%
  \BibitemOpen
  \bibfield  {author} {\bibinfo {author} {\bibfnamefont {R.}~\bibnamefont
  {Meservey}}, \bibinfo {author} {\bibfnamefont {P.~M.}\ \bibnamefont
  {Tedrow}}, \ and\ \bibinfo {author} {\bibfnamefont {R.~C.}\ \bibnamefont
  {Bruno}},\ }\href {http://link.aps.org/doi/10.1103/PhysRevB.11.4224}
  {\bibinfo  {journal} {Phys. Rev. B} \textbf {\bibinfo {volume} {11}},\
  \bibinfo {pages} {4224} (\bibinfo {year} {1975})}\BibitemShut {NoStop}%
\bibitem [{\citenamefont {M\"unzenberg}\ and\ \citenamefont
  {Moodera}(2004)}]{muenzenberg2004}%
  \BibitemOpen
  \bibfield  {author} {\bibinfo {author} {\bibfnamefont {M.}~\bibnamefont
  {M\"unzenberg}}\ and\ \bibinfo {author} {\bibfnamefont {J.~S.}\ \bibnamefont
  {Moodera}},\ }\href {http://link.aps.org/doi/10.1103/PhysRevB.70.060402}
  {\bibinfo  {journal} {Phys. Rev. B} \textbf {\bibinfo {volume} {70}},\
  \bibinfo {pages} {060402} (\bibinfo {year} {2004})}\BibitemShut {NoStop}%
\bibitem [{\citenamefont {Devoret}\ \emph {et~al.}(1990)\citenamefont
  {Devoret}, \citenamefont {Esteve}, \citenamefont {Grabert}, \citenamefont
  {Ingold}, \citenamefont {Pothier},\ and\ \citenamefont
  {Urbina}}]{devoret1990}%
  \BibitemOpen
  \bibfield  {author} {\bibinfo {author} {\bibfnamefont {M.~H.}\ \bibnamefont
  {Devoret}}, \bibinfo {author} {\bibfnamefont {D.}~\bibnamefont {Esteve}},
  \bibinfo {author} {\bibfnamefont {H.}~\bibnamefont {Grabert}}, \bibinfo
  {author} {\bibfnamefont {G.-L.}\ \bibnamefont {Ingold}}, \bibinfo {author}
  {\bibfnamefont {H.}~\bibnamefont {Pothier}}, \ and\ \bibinfo {author}
  {\bibfnamefont {C.}~\bibnamefont {Urbina}},\ }\href
  {http://link.aps.org/abstract/PRL/v64/p1824} {\bibinfo  {journal} {Phys. Rev.
  Lett.} \textbf {\bibinfo {volume} {64}},\ \bibinfo {pages} {1824} (\bibinfo
  {year} {1990})}\BibitemShut {NoStop}%
\bibitem [{\citenamefont {Doma\ifmmode~\acute{n}\else \'{n}\fi{}ski}\ and\
  \citenamefont {Donabidowicz}(2008)}]{domanski2008}%
  \BibitemOpen
  \bibfield  {author} {\bibinfo {author} {\bibfnamefont {T.}~\bibnamefont
  {Doma\ifmmode~\acute{n}\else \'{n}\fi{}ski}}\ and\ \bibinfo {author}
  {\bibfnamefont {A.}~\bibnamefont {Donabidowicz}},\ }\href
  {http://link.aps.org/doi/10.1103/PhysRevB.78.073105} {\bibinfo  {journal}
  {Phys. Rev. B} \textbf {\bibinfo {volume} {78}},\ \bibinfo {pages} {073105}
  (\bibinfo {year} {2008})}\BibitemShut {NoStop}%
\bibitem [{\citenamefont {Koerting}\ \emph {et~al.}(2010)\citenamefont
  {Koerting}, \citenamefont {Andersen}, \citenamefont {Flensberg},\ and\
  \citenamefont {Paaske}}]{koerting2010}%
  \BibitemOpen
  \bibfield  {author} {\bibinfo {author} {\bibfnamefont {V.}~\bibnamefont
  {Koerting}}, \bibinfo {author} {\bibfnamefont {B.~M.}\ \bibnamefont
  {Andersen}}, \bibinfo {author} {\bibfnamefont {K.}~\bibnamefont {Flensberg}},
  \ and\ \bibinfo {author} {\bibfnamefont {J.}~\bibnamefont {Paaske}},\ }\href
  {http://link.aps.org/doi/10.1103/PhysRevB.82.245108} {\bibinfo  {journal}
  {Phys. Rev. B} \textbf {\bibinfo {volume} {82}},\ \bibinfo {pages} {245108}
  (\bibinfo {year} {2010})}\BibitemShut {NoStop}%
\bibitem [{\citenamefont {Franke}\ \emph {et~al.}(2011)\citenamefont {Franke},
  \citenamefont {Schulze},\ and\ \citenamefont {Pascual}}]{franke2011}%
  \BibitemOpen
  \bibfield  {author} {\bibinfo {author} {\bibfnamefont {K.~J.}\ \bibnamefont
  {Franke}}, \bibinfo {author} {\bibfnamefont {G.}~\bibnamefont {Schulze}}, \
  and\ \bibinfo {author} {\bibfnamefont {J.~I.}\ \bibnamefont {Pascual}},\
  }\href {http://www.sciencemag.org/content/332/6032/940.abstract} {\bibinfo
  {journal} {Science} \textbf {\bibinfo {volume} {332}},\ \bibinfo {pages}
  {940} (\bibinfo {year} {2011})}\BibitemShut {NoStop}%
\bibitem [{\citenamefont {Shen}\ and\ \citenamefont {Rowell}(1968)}]{shen1968}%
  \BibitemOpen
  \bibfield  {author} {\bibinfo {author} {\bibfnamefont {L.~Y.~L.}\
  \bibnamefont {Shen}}\ and\ \bibinfo {author} {\bibfnamefont {J.~M.}\
  \bibnamefont {Rowell}},\ }\href
  {http://link.aps.org/doi/10.1103/PhysRev.165.566} {\bibinfo  {journal} {Phys.
  Rev.} \textbf {\bibinfo {volume} {165}},\ \bibinfo {pages} {566} (\bibinfo
  {year} {1968})}\BibitemShut {NoStop}%
\bibitem [{\citenamefont {Wolf}\ and\ \citenamefont {Losee}(1970)}]{wolf1970}%
  \BibitemOpen
  \bibfield  {author} {\bibinfo {author} {\bibfnamefont {E.~L.}\ \bibnamefont
  {Wolf}}\ and\ \bibinfo {author} {\bibfnamefont {D.~L.}\ \bibnamefont
  {Losee}},\ }\href {http://link.aps.org/doi/10.1103/PhysRevB.2.3660} {\bibinfo
   {journal} {Phys. Rev. B} \textbf {\bibinfo {volume} {2}},\ \bibinfo {pages}
  {3660} (\bibinfo {year} {1970})}\BibitemShut {NoStop}%
\end{thebibliography}%

\end{document}